 \documentstyle[12pt]{article}
\if@twoside  m
\else
\fi
 1
 1
\font\sr = msbm10 scaled \magstep 1

\newcommand{\spav}[1]{\parbox{1mm}{\vspace*{#1}}}

\begin{document}
\begin{titlepage}
\begin{flushright}
CERN-TH.7417/94
\end{flushright}
\spav{.1cm}
\begin{center}
{\bf Differential Calculus on Quantum Groups:\\ Constructive Procedure}
\spav{1.3cm}\\
{ B. Jur\v co}\\
{\em CERN, Theory Division},
{\em CH-1211 Geneva 23, Switzerland}

\end{center}
\spav{4cm}
\begin{center}
{\bf Abstract}
\end{center}
{\small
A brief review of the construction and classifiaction of the bicovariant
differential calculi on quantum groups is given.}

\spav{6cm}

Lecture given at the Varenna School on ``Quantum Groups and Their Applications
in Physics"

\spav{1cm}\\
CERN-TH.7417/94\\
August 1994

\end{titlepage}

\newpage

\setcounter{footnote}{0}
\setcounter{page}{1}
{\thispagestyle{empty}}

\section{Preliminaries, notation}

The aim of the present lecture is to give, following the paper \cite{Jurco}, a
procedure for an explicit construction of bicovariant differential calculi on
quantum groups (see also \cite{SW},\cite{Bernard1}). We shall also mention some
recent classification results \cite{SS}. Our interest will lie mainly in the
underlying ideas and we shall omit technical details in our exposition. The
interested reader can consult, for these, the original papers or some nice
review papers \cite{zum}, \cite{LP}.
We shall use the summation rule and the standard notation for the result
of coproduct $\Delta a = \sum a_{(1)}\otimes a_{(2)}$. The antipode will be
denoted by $S$ and the counit by $e$.
We assume that the reader is familiar with the Faddeev--Reshetikhin--Takhtajan
(FRT) \cite{F-R-T} approach to quantum groups and with the general theory of
bicovariant differential calculi on quantum groups due to
Woronowicz \cite{Wordc} (the nice paper \cite{Bernard} may also be helpful in
this connection). In this lecture we shall show a direct relation between these
two constructions.

We shall restrict ourselves to the $q$-deformations of the classical simple
series $A_{n-1}, B_n, C_n$ and $D_n$ as they were introduced in \cite{F-R-T}.
Nevertheless,
as will be easily seen, the procedure works in any case whenever the
corresponding
quantum group and its dual Hopf algebra can be obtained through the FRT
construction \cite{F-R-T}. We shall use its general formulation due to S. Majid
\cite{MajidR}. In particular it means \cite{BM1} that we are starting from some
$R$-matrix ${\cal R}$ ($N^2 \times N^2$--matrix solution of the Yang-Baxter
equation) and we associate a bialgebra $A({\cal R})$ to it. The algebra of
$A({\cal R})$ is generated by the matrix of generators $T=(t_{ij})_{i,j=1}^N$
modulo the relation \cite{F-R-T}
\begin{equation}
{\cal R}T_1T_2=T_2T_1{\cal R}.\label{mt}
\end{equation}
The coalgebra structure is the standard one induced by the matrix
comultiplication.
The notation
\begin{equation}
\Delta T=T\dot\otimes T \label{ct}
\end{equation}
will be used for it.
The counit is given by
\begin{equation}
e(T)=I, \label{ut}
\end{equation}
with $I$ the unit matrix.

Taking a suitable quotient, the bialgebra $A({\cal R})$ is usually made into a
Hopf
algebra, which will be denoted as $A$.
The starting $R$-matrix is assumed to be such, that the Hopf algebra $A$ thus
obtained is a dual-quasitriangular one \cite{ML}. Particularly, we assume that
${\cal R}^{-1}$ and $\tilde {\cal R}=(({\cal R}^{t_2})^{-1})^{t_2}$ exist,
where $t_2$ means the transposition in the second matrix factor, and that
${\cal R}$ can be extended to a functional (denoted for a while by the same
symbol ${\cal R}$) ${\cal R}:A\otimes A \rightarrow \hbox{\sr C}$, such that it
obeys
\begin{equation}
{\cal R}(t_{ij}, t_{kl})={\cal R}_{ik,jl}, \hskip 1cm {\cal R}(St_{ij},
t_{kl})={\cal R}^{-1}_{ik,jl}, \hskip 1cm {\cal R}(t_{ij}, St_{kl})=\tilde
{\cal R}_{ik,jl}
\end{equation}
and
\begin{equation}
{\cal R}(ab,c)= \sum {\cal R}(a,c_{(1)}){\cal R}(a,c_{(2)}), \hskip 1cm {\cal
R}(a,bc)= \sum {\cal R}(a_{(1)},c){\cal R}(a_{(2)},b).
\end{equation}
The standard Jimbo's $R$-matrices \cite{Jimbo} fulfil the above requirements
and the resulting Hopf algebras are the algebras of quantized fuctions
$Fun_q(G)$ of \cite{F-R-T} corresponding to the classical simple series.

The above--mentioned properties of the $R$-matrix can be used in this case to
introduce a
Hopf algebra $U({\cal R})$ dual to the Hopf algebra $Fun_q(G)$ generated by two
matrices of generators $L^{\pm}=(l^{\pm}_{ij})_{i,j=1}^{N}$ of \cite{F-R-T}.
The following commutation relations take place
\begin{equation}
{\cal R}_{21}L^{\pm}_1L^{\pm}_2=L^{\pm}_2L^{\pm}_1{\cal R}_{21},\label{ml}
\end{equation}
$${\cal R}_{21}L^{+}_1L^{-}_2=L^{-}_2L^{+}_1{\cal R}_{21}.$$
The comultiplication on matrices $L^{\pm}$ is again the matrix one
\begin{equation}
\Delta L^{\pm} =L^{\pm}\dot\otimes L^{\pm}\label{cl}
\end{equation}
and the counit is given by
\begin{equation}
e(L^{\pm})=I. \label{ul}
\end{equation}
The pairing between Hopf algebras $Fun_q(G)$ and $U(\cal {R})$ is given on
generators as
\begin{equation}
(L^{\pm},T)={\cal R}^{\pm},\label{pairing}
\end{equation}
where ${\cal R}^+={\cal R}_{21}$ and ${\cal R}^-={\cal R}^{-1}$. For more
details about the matrices  $L^{\pm}$, we refer the reader to the
above--mentioned paper \cite{F-R-T}, which describes also the relation between
the Hopf algebra $U({\cal R})$ and the quantized
enveloping algebra $U_h(g)$ of Drinfeld and Jimbo \cite{Dr}, \cite{Ji}.

Concerning the bicovariant differential calculi on $A=Fun_q(G)$, let us start
with the following fact, proved by Woronowicz \cite{Wordc}. Let us assume that
we are given a family of functionals $F=(f_{ij})_{i,j=1}^{k}\in U\equiv U({\cal
R}),$ $k\in \hbox{\sr N}$, such that
\begin{equation}
\Delta F=F\dot\otimes F\label{cf}
\end{equation}
and
\begin{equation}
e(F)=I\label{uf},
\end{equation}
and a family of quantum functions $R=(R_{ij})_{i,j=1}^k\in A$, such that
\begin{equation}
\Delta R=R\dot\otimes R \label{cr}
\end{equation}
and
\begin{equation}
e(R)=I.\label{ur}
\end{equation}
Besides, matrices $R$ and $F$ are supposed to satisfy the following
compatibility condition
\begin{equation}
R_{ij}(a\ast f_{ih})=(f_{ji}\ast a)R_{hi}, \label{cc}
\end{equation}
for all $j,\,h$ and any $a\in A$. Here we used the notation
$$x\ast a=\sum a^{(1)}(x,a^{(2)}), \hskip 1cm a\ast x=\sum (x,a^{(1)})a^{(2)}$$
for $x\in U$, $a\in A$.
Let us now assume a free left module $\Gamma $ over $A$ generated by elements
$\omega_i,\, i=1,2,...,k$, and let us introduce the right multiplication by
elements of $A$ and the left $\delta_L$ and right coaction $\delta_R$ of $A$ on
$\Gamma$ by the following formulae
\begin{equation}
(a_i\omega_i)b=a_i(f_{ij}\ast b)\omega_j,
\end{equation}
\begin{equation}
\delta_L (a_i\omega_i)= \Delta(a_i)(1\otimes \omega_i),
\end{equation}
\begin{equation}
\delta_R (a_i\omega_i)= \Delta(a_i)(\omega_j \otimes R_{ji}).
\end{equation}
A theorem of Woronowicz says that the triple $(\Gamma, \delta_L, \delta_R)$
is a bicovariant bimodule and, vice versa, that any bicovariant bimodule is of
this form.
Elements $\omega_i,\,i=1,2...,k$ form a basis in the linear subspace
$\,_{inv}\Gamma \subset \Gamma$ of all left-invariant elements of $\Gamma$. We
know from the general theory of Woronowicz that the space $\Gamma$ of all
one-forms on the quantum group $A$ and the whole exterior algebra
$\Gamma^\wedge$ over $A$ are naturally equipped with a structure of a
bicovariant bimodule.

It is an almost obvious fact that having two bicovariant bimodules $\Gamma_1$
and $\Gamma_2$  we can construct their tensor product $\Gamma_1\otimes
\Gamma_2$, which is again a bicovariant bimodule. The linear basis in
$\,_{inv}(\Gamma_1\otimes \Gamma_2)$ can be chosen as
$\omega_{ij}=\omega_i\otimes \omega_j$. In this basis we have
$R_{ij,kl}=R^{1}_{ik}R^2_{jl}$ and $f_{ij,kl}=f^{1}_{ik}f^2_{jl}.$

\section{Construction of differential calculi}

It is enough to consider only the first-order differential calculi on $A$
because we know that a given first-order differential calculus on $A$ can be
uniquely
extended to an exterior differential calculus on $A$ \cite{Wordc}.

Let us now assume the vector corepresentation of $A$ given by the matrix of the
generators $T=(t_{ij})_{i,j=1}^{N}$. A comparison of (\ref{cr}) and (\ref{ur})
with (\ref{ct}) and (\ref{ut})
suggests to try to find a bicovariant bimodule such that $R=T$. It is easily
seen that we can choose both $F=S(L^{\pm})^t$ in this case. The conditions
(\ref{cf}) and (\ref{uf})
are satisfied due to (\ref{cl}) and (\ref{ul}) and the antipode properties. The
compatibility condition (\ref{cc}) is sufficient (because of (\ref{cf})) to
check only for $a=t_{ij}$. In this case, (\ref{cc}) is equivalent to
(\ref{mt}).
The situation is quite similar for the choice $R=S(T)^t$. In this case we can
take $F=L^{\pm}$ by the same reasoning. We denote the bicovariant
bimodules thus obtained as $\Gamma_1$, $\Gamma_2$, $\Gamma_1^c$ and
$\Gamma_2^c$ according to the choices
\begin{equation}
\Gamma_1:\, R=T, \hskip 1cm F=S(L^+)^t,
\end{equation}
\begin{equation}
\Gamma_2:\, R=T, \hskip 1cm F=S(L^-)^t,
\end{equation}
\begin{equation}
\Gamma_1^c:\, R=S(T)^t, \hskip 1cm F=L^-,
\end{equation}
\begin{equation}
\Gamma_2^c:\, R=S(T)^t, \hskip 1cm F=L^+.
\end{equation}
Now we take the tensor product $\Gamma  =  \Gamma_1\otimes \Gamma_1^c$
   to  get  a  new
bicovariant bimodule. For other choices of tensor products $(\Gamma_1^c\otimes
\Gamma_1, \Gamma_2\otimes \Gamma_2^c$ and $\Gamma_2^c\otimes \Gamma_2)$
all that follows  is quite analogous, or it leads to the trivial
differential calculi ($da = 0$, for all $a \in
A$) in the cases $ \Gamma_1\otimes \Gamma_2^c, \Gamma_1^c\otimes \Gamma_2,
\Gamma_2\otimes \Gamma_1^c$ and $\Gamma_2^c\otimes \Gamma_1$.
According to the general theory $\Gamma$ can be described  as  follows.
Let $(\omega_{ij})_{i,j=1}^{N}$ be a  basis  for $ \,_{inv}\Gamma$.
Right multiplication is given by
\begin{equation}
\omega_{ij}a= ((id\otimes S(l^+_{ki})l^-_{jl}  )\Delta a)\omega_{kl}
     \end{equation}
and the right coaction by
\begin{equation}
\delta_R(\omega_{ij})=\omega_{kl}\otimes  t_{ki}S(t_{jl}).\label{action}
\end{equation}
Our choice of the bicovariant bimodule $\Gamma$ is motivated by the particular
form of the coaction (\ref{action}). It follows that
the linear space $\,_{inv}\Gamma$  contains  a  bi-invariant  element
$\tau = \sum \omega_{ii}$, which can be used to define a derivative on $A$.
For $a\in A$  we set
\begin{equation} da = \tau a - a\tau.\label{d}
\end{equation}
It  can  be  easily  checked  that  such a derivative has  all
properties stated in \cite{Wordc}.
The bi-invariance  of  $\tau$ is essential for the differential calculus to be
a bicovariant one.  In  the
case of real forms $Fun(G_q ,\varepsilon_i )$ introduced in \cite{F-R-T}
the *-structure
on $\,_{inv}\Gamma$ is given by
\begin{equation}
\omega_{ij}^* = - \varepsilon_i\varepsilon_j\omega_{ji}.
\end{equation}
According to the rules of \cite{Wordc} the derivative (\ref{d}) and the
*-structure
extend uniquely to the whole $\Gamma^\wedge$. The result for the derivative is
\begin{equation}
d\theta = \tau\wedge\theta - (-1)^k \theta\wedge \tau,
\end{equation}
where $k$ is the degree  of  a  homogeneous  element
$\theta\in \Gamma^\wedge$.

Thus (\ref{d}) defines a *-calculus. In the case of $SU_{q} (2)$, we get  in
this way the $4D_+$  calculus of Woronowicz \cite{P},\cite{PW}.
For $a\in A$ we have
\begin{equation}
          da = ((id\otimes(S(l_{ki}^+)l_{il}^-  -
\delta_{kl}  e)\Delta a)\omega_{kl}  .
\end{equation}
Let us denote by
\begin{equation}
\chi_{ij} = S(l^+_{ik}  )l^-_{ki}  - \delta_{ij}  e ,
\end{equation}
or more compactly
\begin{equation}
\chi = S(L^+)L^- - Ie,
\end{equation}
the matrix of left-invariant vector fields $\chi_{ij}$ on $A$.
The ``commutators" ($[\chi',\chi]=\sum S(\chi_{(1)})\chi'\chi_{(2)}$)  among
the elements $\chi_{ij}$   of the basis dual to the
$\omega_{ij}$ can now
be obtained directly from relations (\ref{ml}) between the
functionals $l^{\pm}_{ij}$ or from
the fact  that $d^2 (a) = 0$ for any $a\in\cal {A}$. We employ the notation
$\lambda_{ijkl,mnop}=(S(l^+_{oi})l^-_{jp})(t_{mk}S(t_{ln}))$
($\lambda$'s are easily expressed with the help of matrices ${\cal R}$ and
$\tilde{\cal R}$).
We have
\begin{equation}
[\chi_{ij},\chi_{kl}]=\chi_{ij}\chi_{kl}-\lambda_{mnop,ijkl}\chi_{mn}\chi_{op}=
-\delta_{kl}\chi_{ij}+\lambda_{ssmn,ijkl}\chi_{mn}.
\end{equation}
In a more compact notation
\begin{equation}
{\cal R}_{21}^{-1}\chi_1{\cal R}_{12}^{-1}\chi_2 - \chi_2{\cal
R}_{21}^{-1}\chi_1{\cal R}_{12}^{-1}=\chi_2{\cal R}_{21}^{-1} {\cal
R}_{12}^{-1}-{\cal R}_{21}^{-1} {\cal R}_{12}^{-1}\chi_2.
\end{equation}
For the real forms the *-structure implies
\begin{equation}
\chi_{ij}^*= \varepsilon_i\varepsilon_j
\chi_{ji}
\end{equation} on left-invariant vector fields $\chi_{ij}$.

Let us note that the set of projections $P(dt_{mn}) =  ((S(l_{ki}^+)l_{il}^-  -
\delta_{kl}  e), t_{mn})\omega_{kl}$ of the differentials of generators
$t_{mn}$ to the space $\,_{inv}\Gamma$ can be chosen
as another basis of $\,_{inv}\Gamma$.

Now we can describe  a  direct  relation  between the enveloping
algebra generated by $(\chi_{ij})_{i,j=1}^N$ and the algebra $U$
of \cite{F-R-T} generated by functionals $(l^{\pm}_{ij})_{i,j=1}^N$.
Let us introduce matrix  $L=(l_{ij})_{i,j=1}^N,$
$L = S(L^+ )L^-.$
A similar matrix $ L^+ S(L^- )$  was introduced in \cite{MR} and  investigated
in \cite{B}. The upper and lower triangular matrices $S(L^+)$ and $L^-$ can
be constructed from  $L$ by  its  decomposition  into
triangular  parts  as  described  in  \cite{B}.  In  this  sense  the
enveloping algebra generated by $\chi$'s and  the  algebra $U$
of FRT
are equivalent.
Let us now discuss very briefly the classical limit. We have
$R = 1 + \hbar r +...,$
where $q = e^\hbar$  and $r$ is the corresponding classical $r$-matrix,
$L^{\pm}  = 1 +  \hbar\eta^{\pm} + ..., $
with $\eta^{\pm}$ matrices of generators for  the  corresponding  Lie
algebra $g$. The matrix elements of $\chi = (\eta^- -\eta^+ )$ are no longer
linearly
independent in this limit, and the linear space spanned by these is
just the Lie algebra $g$. As a result the classical differential calculus on
the group $G$ is obtained as a quotient.

\section{Some remarks on the classification of differential calculi}

Let us mention that for the differential calculi described in the previous
section the dimension of the space of left-invariant forms $\,_{inv} \Gamma$
is $N^2$, where $N$ is the dimension of the vector representation of the
corresponding classical simple group $G$. So it appears higher, as in the
classical case. This is the price we have to pay for the bicovariance of the
corresponding differential calculi. Nevertheless, these differential calculi
are quite natural and they contain the classical differential calculus on the
corresponding group $G$ in the limit. Moreover, the assumption of bicovariance
is a natural one and, besides, it is also technically important. There is no
general theory of left-(right-) invariant differential calculi only.

Another difference, contrary to the classical case, appears. In general there
are many non-isomorphic differential calculi on a given quantum
group, and up to now there is a lack of a functorial method to construct a
natural one.

Two differential calculi $(\Gamma_1,d_1)$ and $(\Gamma_2,d_2)$ are assumed to
be isomorphic if there is a bimodule isomorphism $\phi: \Gamma_1 \rightarrow
\Gamma_2$ and $\phi \circ d_1 = d_2$.
Motivated in the previous section, the following question arises. How many
non-isomorphic bicovariant differential calculi exist, such that the space of
left- invariant forms $\,_{inv}\Gamma$ is spanned by the (not necessary
linearly independent) left-invariant forms
$P(t_{ij})$, there exist
for a given quantum group $A$ corresponding to the one of the classical simple
groups? This was investigated in \cite{SS}, $q$ not being a root of unity
and under the assumption $\mbox{dim} \, _{inv}\Gamma \geq 2$. The discussion
differs for the case of $SL(N)$ and for the remaning classical simple series
$B$, $C$, $D$. The cases $SL(2)$ and $Sp(2)$  should also be treated
separately.

In the $SL(N)$ case the result is obtained by the following generalization of
Section. 1.
There still is an ambiguity in the discussion done there, which amounts to
replacing the standard $R$-matrix for the $A_{N-1}=SL(N)$ case of \cite{F-R-T}
by a new
one, ${\cal R}_p$, which differs by only a factor $p$ (equal to some of the
$N$-th root of unity) from the standard one. This new $R$-matrix can be used in
the FRT construction, which now leads us to new matrices of generators
$L^{\pm}_p$ (the matrix $T$ remains of course the same) and we can repeat the
construction of the differential calculi as above, now starting with the
bicovariant bimodules  $\,_{p_k}\Gamma^+  =  \,_p\Gamma_1\otimes
\,_{p'}\Gamma_1^c$ and $\,_{p_k}\Gamma^-  =  \,_p\Gamma_2\otimes
\,_{p'}\Gamma_2^c$, where $p_k, \,k=0,1,...,N-1$ is such $N$-th root of unity
that $pp'=p_k$. The indices $p$ and $p'$ mean that the matrices $L^{\pm}_p$ or
$L^{\pm}_{p'}$ were used
in the construction of the corresponding bimodules. The resulting tensor
products depend only on $k$.
The rest of the construction is the same as in the Section 1. So we obtain $2N$
differential calculi in the $SL(N)$ case, corresponding to various choices of
the signs $\pm$ and of the integer $k=$.
It can be shown \cite{SS} that, except for a finite number of values of $q$,
this list exhausts all non-equivalent bicovariant differential calculi on
$SL_q(N)$, $N>2$, $q$ not being a root of unity, such that the space of
left-invariant forms $\,_{inv}\Gamma$ is spanned by the left-invariant forms
$P(dt_{ij})$ and $\mbox{dim} \, _{inv}\Gamma \geq 2$. The exceptional values of
$q$ are also discussed in \cite{SS}. Only the two calculi corresponding to
$k=0$ contain the ordinary classical calculus in the limit $q\rightarrow 1$ as
a quotient.
In the remaining case of $SL_q(2)$,
the calculi corresponding to the same $k$ but to different signs $\pm$ can be
easily shown to be isomorphic and the only non-isomorphic calculi are the
$4D_+$ and $4D_-$ calculi of Woronowicz, as was shown in \cite{St} .

A similar discussion can be done also in the remaining cases, corresponding to
the simple classical groups of the types $B_n$, $C_n$, $D_n$ \cite{SS}.
We set $N=2n+1$ in the case $B_n$, and  $N=2n$ in the cases $C_n$ and $D_n$.
To formulate the result we need the differential calculus based on the bimodule
$\Gamma  =  \Gamma_1\otimes \Gamma_1^c$ and also an additional one, which is
described as follows. It is again a tensor product $\Gamma '  =
\Gamma_1\otimes \Gamma_0$,
where the bimodule $\Gamma_0$ is constructed by setting $R=S(T)^t$ and $F=L^0$.
The  matrix of functionals $L^0$ is defined as
$$(L^0,a)=(-1)^k(L^-,a),$$
for $a\in A$ a homogeneous polynomial of order $k$ in the generators $t_{ij}$.
The remaining part of the construction of the corresponding differential
calculus is the same as in Section 1.
Let $N\geq 3$ and assume that $q$ is not a root of unity. Except for finitely
many
values of $q$, the above-described differential calculi are the only
non-isomorhipc bicovariant differential calculi of dimension $N^2$ \cite{SS}.
The missing case of $Sp_q(2)$ is already covered by the discussion of the
$SL_{q^2}(2)$, which is isomorphic to $Sp_q(2)$.
Again, only the calculus $(\Gamma, d)$ contains the ordinary classical calculus
in the limit $q\rightarrow 1$ as a quotient.

\section{Remarks}
Let us mention other examples of quantum groups for which the construction of
Section 1 works. The examples include the quantum group $GL_q(n)$ and its
more-parametric modifications (see \cite
{leon1}, \cite{Sun-Wang},\cite{Malt}, \cite{MH},\cite{SCH}, \cite{LP}, and many
others), "
``complex" quantum groups (real forms of the dual to the Drinfeld's quantum
double) corresponding to the classical simple groups \cite {D2}, \cite{D3},
\cite{Satoshi}, the inhomogeneous quantum groups
\cite {leon},\cite{W},\cite{leon2}, \cite{LP}, etc.
For applications and subsequent developement, let us mention for instance the
$q$-deformed BRST complex \cite{Leon3}, \cite{Sat}, the $q$-deformed standard
complex \cite{Bernie}, the representation theory of quantized enveloping
algebras \cite{CS}, quantum mechanics on quantum spaces \cite{Munich}, the
$q$-deformd gravity \cite{leon}, the Cartan calculus on quantum groups
\cite{Schupp}, \cite{Paolo},  etc.
For alternative approaches, see e.g. \cite{Zumino}, \cite{BM1}, \cite{Malt},
\cite{Fadd}, \cite{Manin}, \cite{BraidMajid}, \cite{Sud}.
It is almost impossible to mention all the papers related to the subject and we
apologize to those authors whose work has been omitted.

\end{document}